\documentclass[10pt,preprint]{aastex}
 \usepackage{natbib}
\def\ang{\AA}
\def\arcsec{\hbox{$^{\prime\prime}$}}

\def\gapprox{\lower.4ex\hbox{$\;\buildrel >\over{\scriptstyle\sim}\;$}}
\def\lapprox{\lower.4ex\hbox{$\;\buildrel <\over{\scriptstyle\sim}\;$}}

\shortauthors{Aschwanden 2022}
\shorttitle{IRIS Observations of Fractal Dimension}

\begin{document}
\renewcommand{\topfraction}{0.95}
\renewcommand{\bottomfraction}{0.95}
\renewcommand{\textfraction}{0.05}
\renewcommand{\floatpagefraction}{0.95}
\renewcommand{\dbltopfraction}{0.95}
\renewcommand{\dblfloatpagefraction}{0.95}


\title{ Interface Region Imaging Spectrograph (IRIS) Observations of 
	the Fractal Dimension in the Solar Atmosphere }

\author{Markus J. Aschwanden}
\affil{Lockheed Martin, Solar and Astrophysics Laboratory (LMSAL),
       Advanced Technology Center (ATC),
       A021S, Bldg.252, 3251 Hanover St.,
       Palo Alto, CA 94304, USA;
       e-mail: aschwanden@lmsal.com}

\and 
\author{Nived Vilangot Nhalil}
\affil{Armagh Observatory and Planetarium, College Hill, Armagh BT61 9DG, UK}

\begin{abstract}
While previous work explored the fractality and self-organized criticality
(SOC) of flares and nanoflares in wavelengths emitted in the solar corona (such
as in hard X-rays, soft X-rays, and EUV wavelenghts), we focus here on
impulsive phenomena in the photosphere and transition region, as observed with the 
{\sl Interface Region Imaging Spectrograph (IRIS)} in the temperature range of
$T_e \approx 10^4-10^6$ K. We find the following fractal dimensions (in increasing
order): $D_A=1.21 \pm 0.07$ for photospheric granulation,
$D_A=1.29 \pm 0.15$ for plages in the transition region,
$D_A=1.54 \pm 0.16$ for sunspots in the transition region,
$D_A=1.59 \pm 0.08$ for magnetograms in active regions,
$D_A=1.56 \pm 0.08$ for EUV nanoflares,
$D_A=1.76 \pm 0.14$ for large solar flares, and up to
$D_A=1.89 \pm 0.05$ for the largest X-class flares.
We interpret low values of the fractal dimension 
($1.0 \lapprox D_A \lapprox 1.5$)
in terms of sparse curvi-linear flow patterns, while high values
of the fractal dimension ($1.5 \lapprox D_A \lapprox 2.0$) 
indicate near space-filling transport processes, such as
chromospheric evaporation. Phenomena in the solar transition
region appear to be consistent with SOC models, based on 
their size distributions of fractal areas $A$ and (radiative) energies $E$, 
which show power law slopes of $\alpha_A^{obs}=2.51 \pm 0.21$
(with $\alpha_A^{theo}=2.33$ predicted), and $\alpha_E^{obs}=2.03 \pm 0.18$
(with $\alpha_E^{theo}=1.80$ predicted).
\end{abstract}


\section{	INTRODUCTION 		}  

There are at least three different approaches that quantify
the statistics of nonlinear processes with the concept of
{\sl self-organized criticality (SOC)} and fractality: 
(i) microscopic models, (ii) macroscopic models, and (iii) 
observations of power laws and scaling laws. The microscopic 
SOC models consist of numerically simulated avalanches that evolve
via next-neighbor interactions in a lattice grid (Bak et 
al.~1987, 1988), also called {\sl cellular automatons},
which have been quantized up to numerical limits of 
$\approx 10^6-10^9$
cells per avalanche process. The macroscopic models describe
the nonlinear evolution of (avalanching) instabilities
with analytical (geometric and energetic) quantities, which
predict physical scaling laws and power law-like occurrence
frequency size distributions. The third category of SOC approaches 
includes observations with fitting of power law-like
distribution functions and waiting time distributions,
which provide powerful tests of theoretical SOC models.
A total of over 1500 SOC-related publications appeared
at the time of writing. For brevity, we mention a few
textbooks only (Bak 1996; Aschwanden 2011; Pruessner 2012),
and a recent collection of astrophysical SOC reviews, 
presented in the special volume {\sl Space Science Reviews}
Vol. 198 (Watkins et al.~2016; Aschwanden et al.~2016;
Sharma et al.~2016; McAteer et al.~2016).  

In this paper we focus on SOC modeling of impulsive events
detected in the solar atmosphere,
as observed with the {\sl Interface Region Imaging
Spectrograph (IRIS)} (De Pontieu et al.~2014), while
solar flare events observed in hard X-rays, soft X-rays,
and {\sl Extreme-Ultraviolet (EUV)} have been compared in
recent studies (Aschwanden 2022a, 2022b). Large solar flares
observed in hard and soft X-rays show typically electron 
temperatures of $T_e \approx 5-35$ K, while coronal
nanoflares observed in EUV have moderate temperatures
of $T_e \approx 1-2$ MK. Hence it is interesting to
investigate transition region events, which are observed in a 
different temperature regime ($T_e \approx 10^4-10^6$ K) 
than coronal phenomena. In a previous study with the same
IRIS data, it was found that plages and sunspots have
different power law indices for the areas of events, 
being smaller ($\alpha_A 
\lapprox 2$) for sunspot-dominated active regions, 
and larger for plage regions, $\alpha \gapprox 2$,
(Nhalil et al.~2020).
If both coronal and transition region brightenings exhibit
the same SOC behavior and are produced by the same
physical mechanism, one would expect the same fractal 
dimension and power law slope of the occurrence 
frequency size distribution, which is an important 
test of the coronal heating problem. 

The structure of this paper consists of an observational
Section 2, a theoretical modeling Section 3, a discussion
Section 4, and a conclusion Section 5. 

\section{		OBSERVATIONS 	 		}

This is a follow-on study of previous work, 
``The power-law energy distributions of small-scale impulsive 
events on the active Sun: Results from IRIS'' (Nhalil et al.~2020). 
We call these small-scale impulsive events simply ``events'', 
which possibly could be related to ``nanoflares'' or ``brightenings''.
In the previous study, 12 IRIS datasets were investigated with
an automated pattern recognition algorithm, yielding statistics 
of three parameters, namely the event area $A$ (in units of pixels),
the event (radiative) energy $E$ (in units of erg), and event durations
or lifetimes $T$ (in units in seconds). IRIS has pixels with
a size of $0.17\arcsec \approx 0.123$ Mm, which have been rebinned
to $L_{pixel}=0.33\arcsec \approx 0.247$ Mm. The pixel size of areas 
thus corresponds to $A_{pixel}=L_{pixel}^2=0.247^2$ Mm$^2$= 
0.06076 Mm$^2$. The range of event areas covers $A=4-677$ pixels,
which amounts to length scales of $L=\sqrt{A}=(2-26)$ pixels, or 
$L=(2-26)*0.247$ Mm $\approx$ (0.5-6.4) Mm = (500-6400) km.
The date of observations,
the field-of-view (FOV), the cadence, and the NOAA active region
numbers are given in Table 1 of Nhalil et al.~(2020), for each
of the 12 IRIS datasets.

The automated pattern recognition code was run with different
threshold levels of 3, 5, and 7 $\sigma$ for the detection of
events, from which we use the 3-$\sigma$ level here.
We use Slitjaw images (SJI) of the 1400 \ang\ channel of IRIS,
which are dominated by the Si IV 1394 \ang\ and 1403 \ang\
resonance lines, formed in the transition region. 
Nhalil et al.~(2020) compared also images from the SJI 1330 \ang\
channel, which is dominated by the C II 1335 \ang\ and 1336 \ang\
lines, originating in the upper chromosphere and transition region 
at formation temperatures of $T_e \approx 3 \times 10^4$ K and
$T_e \approx 8 \times 10^4$ K (Rathore and Carlsson 2015;
Rathore et al.~2015).

\subsection{	Size Distributions		}

Our first measurement is the fitting of a power law distribution
function $N(A) \propto A^{-\alpha_A}$ 
of the event (or nanoflare) areas $A$,
separately for each of the 12 IRIS datasets, as shown in Fig.~1.
The area of the event is a combination of all the spatially 
connected 3-$\sigma$ pixels throughout its lifetime.
The lowest bin was discarded when a visible deviation from 
a power law was apparent in the histogram. The number of events
amounts to 23,633 for all 12 datasets together, varying from
65 to 4725 events per IRIS dataset (Table 1). The 
power law slope fits vary from the lowest value $\alpha_A=2.14$ 
(dataset 1) to the highest value $\alpha_A=2.83$ (dataset 4),
having a mean and standard deviation of (Fig.~2, top panel).
\begin{equation}
	a_A^{obs} = 2.51\pm0.21 \ .
\end{equation}
The area size distributions are shown superimposed for the
12 IRIS datasets (Fig.~2, top panel), which illustrates almost identical
power law slopes in different IRIS datasets.

Fitting the energy size distributions, 
$N(E) \propto E^{-\alpha_E}$,
yields the following mean for all 12 IRIS datasets (Fig.~2, middle panel), 
\begin{equation}
	a_E^{obs} = 2.03\pm0.18 \ .
\end{equation}

Fitting the duration size distributions,
$N(T) \propto T^{-\alpha_T}$,
yields the following mean for all 12 IRIS datasets (Fig.~2, bottom panel), 
\begin{equation}
	a_T^{obs} = 2.65\pm0.39 \ .
\end{equation}
We will interpret these power law slopes in terms of SOC models
in Section 4.6.

\subsection{	Fractal Dimension with Box-Counting Method	}

The next parameter that we are interested in is the fractal dimension.
A standard method to determine the fractal dimension $D_A$ of an 
image is the box-counting method, which is defined by the 
asymptotic ($L \mapsto 0$) ratio of the fractal area $A$ to the
the length scale $L$, i.e., $D_A = {\log(A) / \log (L)}$,
also called Hausdorff (fractal) dimension.
We calculate the so-defined Hausdorff dimension $D_A$ for each
of the 12 IRIS datasets (column $D_{A2}$ in Table 1), which
reveals a very narrow spread of values for the fractal dimension, 
with a mean and stadard deviation of $\lapprox 2\%$, 
\begin{equation}
	D_{A2}^{obs} = 1.58\pm0.03 \ .
\end{equation}

The fractal nature of the 12 IRIS datasets is rendered in Figs.~(3)
and (4), where the white areas correspond to zones with enhanced emission,
and black areas correspond to the background with weak emission.
The successive reduction of spatial resolution is shown in Fig.~4 for
$N_{bin}=128, 64, 32, 16$ pixels,
which all converge to the same fractal dimension of $D_A=1.33$. 
An example of a
theoretical fractal pattern with a close ressemblance to the
observed transition region patterns of dataset 8 
is shown in Fig.~5, which is
called the {\sl ``golden dragon fractal''} and has a Hausdorff
dimension of $D_A = 1.61803$. 
 
\subsection{    Comparison of Photospheric, Transition Region, and Coronal 
		Fractal Dimensions }

In Table 3 we compile fractal dimensions obtained from photospheric and
transition region fractal features, which may be different from coronal
and flare-like size distributions. The fractal dimension has been measured
in white-light wavelengths with the perimeter-area method, containing
dominantly granules and supergranulation features (Roudier and Muller 1986;
Hirzberger et al.~1997; Bovelet and Wiehr 2001; Paniveni et al.~2010),
which exhibit a mean value of (Table 3),
\begin{equation}
        D_A^{gran} = 1.21 \pm 0.07 \ ,
\end{equation}
We have to be aware that white-light emission originates in the solar
photosphere, which has a lower altitude than any transition region or coronal
feature. The relatively low value obtained for granulation features thus
indicates that the granulation features seen in optical wavelengths are
almost curvi-linear (with little area-filling geometries), which is expected
for sparse photospheric mass flows along curvi-linear flow lines.

A second feature we consider are plages, measured in magnetograms with the
{\sl linear-area (LA)} method (Balke et al.~1993), and in transition region 
IRIS 1400 \ang\ data (Nhalil et al.~2020), which have formation temperatures 
of $\approx 10^{3.7} - 10^{5.2}$ K in the lower transition region, 
exhibiting a mean value of (Table 3),
\begin{equation}
        D_A^{plage} = 1.26 \pm 0.16 \ .
\end{equation}
This set of IRIS measurements exhibit a relatively low value for the
fractal dimension, similar to the photospheric granulation features.
Based on this low fractal dimension, photospheric flows appear to be
organized along curvi-linear features, rather than solid-area geometries.

A third feature that we investigate are sunspots (Nhalil et al.~2020,
and this work), which reveal higher values of fractal dimension, namely
(Table 3),
\begin{equation}
        D_A^{sunspot} = 1.54 \pm 0.16 \ .
\end{equation}
Apparently, sunspots organize fractal features into space-filling geometries,
where fragmentation into smaller and smaller fractal features is suppressed,
because of the strong magnetic fields that control the penumbral flows of 
sunspots.

A fourth feature is an active region, observed in photospheric
magnetograms and analyzed with the linear-area method
(Lawrence 1991; Lawrence and Schrijver 1993; Meunier 1999, 2004;
Janssen et al.~2003; Ioshpa et al.~2008),
or with the box-counting method (McAteer et al.~2005).
The mean value of fractal dimensions measured in active regions 
is found to be (Table 3),
\begin{equation}
        D_A^{AR} = 1.59 \pm 0.20 \ .
\end{equation}
Apparently, active regions organize magnetic features into space-filling,
area-like geometries, similar to sunspot features. 

A fifth phenomenon is a nanoflare event, which has been
related to the SOC interpretation since Lu and Hamilton (1991). 
Nanoflares have been observed in EUV 171 \ang\ and 195 \ang\ with
the TRACE instrument, as well as in soft X-rays using the 
Yohkoh/SXT (Solar X-Ray Telescope) (Aschwanden and Parnell 2002),
which show a mean value of (see Table 3), 
\begin{equation}
        D_A^{nano} = 1.56 \pm 0.08 \ .
\end{equation}
Nanoflares have been observed in the Quiet Sun and appear to have a
similar fractal dimension as impulsive brightenings in active regions,
as measured in magnetograms. 

For completeness we list also the fractal dimension measured in
large solar flares, for M-class flares, X-class flares, and the
Bastille Day flare (Aschwanden and Aschwanden 2008a), which all
together exhibit a mean value of (Table 3),
\begin{equation}
        D_A^{flare} = 1.76 \pm 0.14 \ .
\end{equation}
This is the largest mean value of any measured fractal dimension,
which indicates that the flare process fills the flare area
almost completely, due to the superposition of many coronal
postflare loops that become filled as a consequence of the
chromospheric evaporation process.

Thus, we can distinguish
at least three groups with significantly different fractal properties 
in photospheric, transition region, and coronal data. A first group
has curvi-linear features in the granulation and in plage features, which
have a relatively low fractal dimension $D_A^{gran} \approx D_A^{plage} 
\approx 1.2-1.3$. There is a second group of sunspot, active region,
and nanoflare phenomena, wich exhibit an intermediate range of 
fractal dimensions of $D_A^{sunspot} \approx D_A^{AR} 
\approx D_A^{nano} \approx 1.5-1.6$.
And there is a third group of large flares (M- and X-class), which
have a fractal dimensions of $D_A^{flare} \approx 1.6-1.9$.

\section{	THEORETICAL MODELING  	 		}

\subsection{	The Hausdorff Fractal Dimension		}

The definition of the fractal dimension $D_A$ for 2-D areas $A$ 
is also called the {\sl Hausdorff dimension} $D_A$ (Mandelbrot 1977),
\begin{equation}
        A = L^{D_A} \ ,
\end{equation}
or explicitly,
\begin{equation} 
	D_A = {\log(A) \over \log (L)} \ ,
\end{equation}
where the area $A$ is the sum of all image pixels $I(x,y) \ge I_0$
above a background threshold $I_0$, and $L$ is the length scale 
of a fractal area. A structure is fractal, when the ratio $D_A$
is approximately constant versus different length scales and
converges to a constant for the smallest length scales $L \mapsto 0$.
The method described here is also called the box-counting method, because
the pixels are counted for the area $A$ and the length scale $L$.

In analogy, a fractal dimension can also be defined for the 3-D volume $V$, 
\begin{equation}
        V = L^{D_V} \ ,
\end{equation}
or explicitly
\begin{equation} 
	D_V = {\log(V) \over \log (L)} \ ,
\end{equation}
The valid range for area fractal dimensions is 
$1 \le D_A \le 2$ and $2 \le D_V \le 3$, where $D=1,2,3$ are the
Euclidean dimensions.

We can estimate the numerical values of the fractal dimensions
$D_A$ and $D_V$ from the means of the minimum and maximum values
in each Euclidean domain,
\begin{equation}
        D_A = {(D_{A,min} + D_{A,max}) \over 2} 
	= {(1 + 2)\over 2} = {3 \over 2} = 1.50 \ ,
\end{equation}
and correspondingly,
\begin{equation}
        D_V = {(D_{V,min} + D_{V,max}) \over 2} 
	= {(2 + 3)\over 2} = {5 \over 2} = 2.50 \ .
\end{equation}
The 2-D fractal dimension $D_A$ is the easiest accessible SOC parameter,
while the 3-D fractal dimension $D_V$ requires information of
fractal structures along the line-of-sight, either using a geometric
or tomographic model, or modeling of optically-thin plasma (in the case of
an astrophysical object observed in soft X-ray or EUV wavelengths).

We find that the theoretical prediction of $D_A=(3/2)=1.50$ (Eq.~15) for the fractal 
area parameter $A$ is approximately consistent with the observed values obtained
with the box-counting method, $D_{A2}^{obs} = 1.58\pm0.03$ (Table 1). 

\subsection{	The SOC-Inferred Fractal Dimension	}

The size distribution $N(L)$ of length scales $L$,
also called the {\sl scale-free probability conjecture}
is (Aschwanden 2012),
\begin{equation}
        N(L) \ dL \propto L^{-d} \ dL \ ,
\end{equation}
where $d$ is the Euclidean space dimension, generally 
set to $d=3$ for most real-world data. Note, that this
occurrence frequency distribution function is simply
a power law, which results from the reciprocal relationship
of the number of events $N(L)$ and the length scale $L$.
Since the fractal dimension $D_A$ for event areas $A$ is defined as
(Eq.~11), 
\begin{equation}
	A = L^{D_A} \ ,
\end{equation}
we obtain the inverse function $L(A)$ ,
\begin{equation}
	L = A^{(1/D_A)} \ ,
\end{equation}
and the derivative, 
\begin{equation}
	\left( {dL \over dA} \right) = A^{(1/D_A-1}) \ ,
\end{equation}
so that we obtain the area distribution $N(A)$ by substituting
of $L$ (Eq.~19) and $dL/dA$ (Eq.~20) into $N(L)$ (Eq.~17),
\begin{equation}
        N(A) \ dA = N[L(A)] \left({dL \over dA}\right) dA 
        =[L(A)]^{-d} \ A^{(1/D_A-1)}\ dA 
	= A^{(-\alpha_A)} \ dA\ ,
\end{equation}
which yields the power law index $\alpha_A$,
\begin{equation}
	\alpha_A = 1 + {(d-1) \over D_A} = 1 + {2 \over D_A} \ .
\end{equation}
Vice versa we can then obtain the SOC fractal dimension $D_A$
from an observed power law slope $\alpha_A$,
\begin{equation}
	D_A = {2 \over (\alpha_A - 1)} \ .
\end{equation}
This is an alternative method (Eq.~23) to calculate the fractal 
area dimension, in contrast to the box-counting method (Eq.~12),
which we call the SOC-inferred fractal dimension, because it uses the
size distribution of areas that are defined in SOC models.
The so calculated fractal dimension $D_{A1}$
exhibits a mean and standard deviation of (Table 1), 
\begin{equation}
        D_{A1} = {2 \over (\alpha_{A1} - 1 ) }
               = 1.35 \pm 0.19 \ ,
\end{equation}
as tabulated for each IRIS dataset in Table 1.

However, there is a significant difference
between the two methods, which is apparent in terms of a much smaller 
spread of values ($\approx 2\%$) for the box-counting method, compared
with the much wider spread of values ($\approx 14\%$) for the power law
fit method. Obviously, the power law fit method is more sensitive to the
spatial variation of individual fractal features than the box-counting method,
while the latter method averages the fractal features, so that
the mean value of the fractal dimension is more robust. 

\section{	DISCUSSION 		}

\subsection{	Basic Fractal Dimension Measurement Methods		}

A fractal geometry is a ratio that provides a statistical index of
complexity, and changes as a function of a length scale that is
used as a yardstick to measure it (Mandelbrot 1977). Well-known 
geometries are the Euclidean dimensions of curves or lines $L$,
being one-dimensional structures ($d=1$), areas $A \propto L^2$,
being two-dimensional structures ($d=2$), and volumes $V = L^3$,
being three-dimensional structures ($d=3$). All other values 
between 1 and 3 are non-integer dimensions and are called fractal 
dimensions. 

Basic methods to measure fractal dimensions include the {\sl linear-area
(LA)} method, the {\sl perimeter-area (PA)} method, and the 
{box-counting (BC)} method. 
The LA method calculates the ratio of a fractal area $A$ to
a space-filled encompassing rectangular area with size $L^2$.
Similarly, the PA method yields a ratio of 
the encompassing curve length or perimeter length ($P=\pi r$ 
in the case of a circular boundary). The box-counting method uses a
cartesian (2-D or 3-D) lattice grid $[x,y]$ and counts all pixels
above some threshold or background, and takes the ratio to
the total counts of all pixels inside the encompassing coordinate grid.
These three methods appear to be very simple, but are not unique.
The resulting fractal dimensions may depend on the assumed
level of background subtraction. The encompassing perimeter
depends on the definition of the perimeter (square, circle,
polygon, etc.). Multiple different geometric patterns may
cause a variation of the fractal dimension across an image or
data cube. Temporal variability can modulate the fractal dimension
as a function of time. Theoretical values of fractal dimensions
converge by definition to a unique value (e.g., $D_A=1.61803$
for the golden dragon fractal, Fig.~5), while observed data
almost always exhibit some spatial inhomogeneity that gives
rise to a spread of fractal dimension values across an image.

\subsection{	Granulation in Photosphere		}

A compilation of fractal dimensions measured in photospheric 
white-light wavelengths (or in magnetograms) is given in Table 3.
The solar granulation has a typical spatial scale of $L=1000$ km, or a
perimeter of $P=\pi L \approx 3000$ km.
Roudier and Muller (1986)
measured the areas $A$ and perimeters $P$ of 315 granules and found
a power law relation $P \propto A^{D/2}$,
with $D=1.25$ for small granules (with perimeters of $P \approx 500-4500$ km)
and $D=2.15$ for large granules (with $P=4500-15,000$ km).
The smaller granules were interpreted in terms of turbulent origin, because
the predicted fractal dimension of an isobaric atmosphere with isotropic
and homogeneous turbulence is $D=4/3\approx 1.33$ (Mandelbrot 1977).
Similar values ($D_A=1.30$ and $D_A=1.16$) were found by Hirzberger et al.~(1997). 
Bovelet and Wiehr (2001) tested different pattern recognition algorithms
(Fourier-based recognition technique FBR and multiple-level tracking MLT)
and found that the value of the fractal dimension strongly depends on
the measurement method. The MLT method yielded a fractal dimension of
$D_A=1.09$, independent of the spatial resolution, the heliocentric
angle, and the definition in terms of temperature or velocity.
Paniveni et al.~(2005) found a fractal dimension of $D_A \approx 1.25$
and concluded, by relating it to the variations of kinetic energy, 
temperature, and pressure, that the supergranular network is close to
being isobaric and possibly of turbulent origin. Paniveni et al.~(2010)
investigated supergranular cells and found a fractal dimension of
$D_A=1.12$ for active region cells, and $D_A=1.25$ for quiet region cells, 
a difference that they attributed to the inhibiting effect of the
stronger magnetic field in active regions. Averaging all fractal 
dimensions related to granular datasets we obtain a mean value of
$D_A=1.21\pm0.07$, which is closer to a curvi-linear topology 
($D_A \gapprox 1.0$) than to an area-filled geometry 
($D_A \lapprox 2.0$).

The physical understanding of solar (or stellar) granulation 
has been advanced by numerical magneto-convection models and 
N-body dynamic simulations, which predict the evolution of 
small-scale (granules) into large-scale
features (meso or supergranulation), organized by surface
flows that sweep up small-scale structures and form clusters of
recurrent and stable granular features (Hathaway et al.~2000;
Berrilli et al.~2005; Rieutord et al.~2008, 2010).
The fractal structure of the solar granulation is obviously a
self-organizing pattern that is created by a combination of
subphotospheric magneto-convection and surface flows, which are
turbulence-type phenomena.

\subsection{	Transition Region 		}

Measurements of the fractal dimension in the transition region
have been accomplished with IRIS 1400 \ang\ observations of plages
and sunspot regions (Nhalil et al.~2020; and this work, see Table 3). 
Fractal dimensions of transition region features were evaluated
with a box-counting method here, yielding a range of $D_A\approx 
1.26\pm 0.13$ for the 8 datasets of plages in the transition region
listed in Table 3,
and $D_A\approx 1.54\pm 0.16$ for the 4 datasets with sunspots
listed in Table 3), following the same identification as used
in the previous work (Nhalil et al.~2020). The structures 
observed in the 1400 \ang\ channel of IRIS are dominated by the 
Si IV 1394 \ang\ and 1403 \ang\ resonance lines, which are formed 
in the transition region temperature range of $T=10^{4.5}-10^6$ K, 
sandwiched between the cooler chromosphere and the hotter corona. 

One prominent feature in the transition region is the phenomenon
of {\sl ``moss''}, which appeas as a bright, dynamic pattern with
dark inclusions, on spatial scales of $L \approx 1-3$ Mm, which
has been interpreted as the upper transition region above active
region plage and below relatively hot loops (De Pontieu et al.~1999).
Besides transition region features, measurements in chromospheric 
(Quiet-Sun) network structures in the temperature range of 
$T=10^{4.5}-10^6$ K yield fractal dimensions of $D_A=1.30-1.70$
(Gallagher et al.~1998). Furthermore, a value of $D_A\approx 1.4$ 
was found for so-called
{\sl Ellerman bombs} (Georgoulis et al.~2002), which are short-lived
brightenings seen in the wings of the H$\alpha$ line from the low
chromosphere. In addition, a range of $D_A\approx 1.25-1.45$ 
was measured from a large
survey of 9342 active region magnetograms (McAteer et al.~2005).
Measurements of SOHO/CDS in EUV lines in the temperature range of
$T_e \approx 10^{4.5}-10^6$ revealed a distinct temperature dependence:
fractal dimensions of $D_A \approx 1.5-1.6$ were identified
in He I, He II, OIII, 
OIV, OV, Ne VI lines at $\log(T_e) \approx 5.8$, then a peak with
$D_A \approx 1.6-1.7$ at $\log(T_e) \approx 5.9$, and a drop of 
$D_A \approx 1.3-1.35$ at $\log(T_e) \approx 6.0$ (see Fig.~11 in
Gallagher et al.~1998).  
The temperature dependence of the fractal dimension can be interpreted
in terms of sparse heating that produces curvi-linear flow patterns
with a low fractal dimensions of $D_A \lapprox 1.5$, while strong 
heating produces volume-filling by chromospheric evaporation 
with high fractal dimensions $D_A \gapprox 1.5$. 

In recent work it was found that the concept of mono-fractals
has to be generalized to multi-fractals to quantify the
spatial structure of solar magnetograms more accurately
(Lawrence et al.~1993, 1996; Cadavid et al.~1994; McAteer et al.~2005;
Conlon et al.~2008).

\subsection{	Photospheric Magnetic Field in Active Regions		}

A number of studies investigated the fractal dimension of the 
photospheric magnetic field, as observed in magnetograms in the
Fe I (6302 \ang , 5250 \ang ) or Ni I (6768 \ang ) lines.
Meunier (1999) evaluated the fractal dimension with the perimeter-area
method and found $D_A=1.48$ for supergranular structures to $D_A=1.68$ for
the largest structures, while the linear size-area method yielded
$D_A=1.78$ and $D_A=1.94$, respectively. In addition, a solar cycle dependence
was found by Meunier (2004), with the fractal dimension varying from
$D_A=1.09\pm0.11$ (minimum) to $D_A=1.73\pm0.01$ for weak-field regions
($B_m<900$ G), and $D_A=1.53\pm0.06$ (minimum) to $D_A=1.80\pm0.01$ for
strong-field regions ($B_m>900$ G), respectively.
A fractal dimension of $D_A=1.41\pm0.05$ was found by
Janssen et al.~(2003), but the value varies as a function of
the center-to-limb angle and is different for a speckle-reconstructed
image that eliminates seeing and noise.

A completely different approach to measure the fractal dimension $D$ was
pursued in terms of a 2-D diffusion process, finding fractal diffusion
with dimensions in the range of $D\approx 1.3-1.8$
(Lawrence 1991) or $D=1.56\pm0.08$ (Lawrence and Schrijver 1993)
by measuring the dependence of the mean square displacement of magnetic
elements as a function of time. Similar results were found by
Balke et al.~(1993). The results exclude Euclidean 2-D diffusion 
but are consistent with percolation theory for diffusion of clusters 
at a density below the percolation threshold 
(Lawrence and Schrijver 1993; Balke et al.~1993).

\subsection{	Coronal Flares 			}

Although this study is focused on the fractal geometry of
transition region features observed 
with IRIS, we compare these results also with coronal values.
The fractal dimension of coronal events has been measured
for 10 X-class flares, 10 M-class flares, and the
Bastille-Day flare (Aschwanden and Aschwanden 2008a, 2008b).
Interestingly, these datasets exhibit relatively large
values of the fractal dimension, with a mean and standard
deviation of $D_A=1.76 \pm 0.14$. They show a trend that the
largest flares, especially X-class flares, exhibit the highest
values of $D_A \lapprox 1.8-1.9$ (Table 3). If we
attribute flare events to the magnetic reconnection
process, the observations imply that the flare plasma
fills up the flare volume with a high space-filling factor,
which is consistent with the chromospheric evaporation process..

Phenomena of smaller magnitude than large flares include
microflares, nanoflares, coronal EUV brightenings, etc.
Such small-scale variability events are found to have
a mean fractal dimension of $D_A=1.56\pm0.08$ (Table 3),
which is compatible with those found in M-class flares,
but clearly has a lower fractal dimension than large
flares, i.e., $D_A=1.76\pm0.14$ (Table 3). 

\subsection{	Self-Organization and Criticality	}

The generation of magnetic structures that bubble up from the
solar convection zone to the solar surface by buoyancy, observed
as emerging flux phenomena in form of active regions, sunspots,
and pores, can be statistically described as a random process,
a self-organization (SO) process, self-organized criticality (SOC),
percolation, or a diffusion process. Random processes produce 
incoherent structures, in contrast to the coherent magnetic
flux concentrations observed in sunspots. 
A self-organization (SO) process needs
a driving force and a counter-acting feedback mechanism that
produces ordered structures (such as the convective granulation cells;
Aschwanden et al.~2018).
A SOC process exhibits power law size distributions of avalanche sizes
and durations. The finding of a fractal dimension of a power law
size distribution in magnetic features
alone is not a sufficient condition to prove or rule out any of these
processes.  Nevertheless, the fractal dimension yields a scaling law 
between areas ($A
\propto L^{D_2}$) or volumes ($V \propto L^{D_3}$), and length scales $L$
that quantifies scale-free (fractal) processes in form of power laws
and can straightforwardly be incorporated in SOC-like models.

If we compare the standard SOC parameters measured in observations
(Fig.~2) with the theoretically expected values from the
standard SOC model (Table 2), we find that the power law slopes
for event areas $A$ agree well 
($a_A^{obs}=2.51\pm0.21$) versus $a_A^{theo}=2.33$ (Fig.~2),
while the power law slopes for the radiated energy $E$ agree within'
the stated uncertainties,
($a_E^{obs}=2.03\pm0.18$) versus $a_E^{theo}=1.80$ (Fig.~2),
but the power law slopes for the time duration $T$ disagree 
($a_T^{obs}=2.65\pm0.39$) versus $a_T^{theo}=2.00$ (Fig.~2).
The latter disagreement is possibly caused by the restriction
of a constant minimum event lifetime (either 60 s or 110 s)
that was assumed in the previous work 
(Nhalil et al.~2020). The interpretation of these results
implies that transition region brightenings have a similar
statistics as the SOC model, at least for active regions,
nanoflares, and large flares, with a typical fractal dimension
of $D_A \approx 1.5-1.6$, but are significantly lower for
photospheric granulation and transition region plages
($D_A \approx 1.2-1.3$), which implies the dominance of
sparse quasi-linear flow structures in the photosphere and
transition region.

\section{	CONCLUSIONS	}

Our aim is to obtain an improved undestanding of fractal 
dimensions and size distributions observed in the solar 
photosphere and transition region, which complement previous
measurements of coronal phenomena, from nanoflares to 
the largest solar flares. Building on the previous study
{\sl ``Power-law energy distributions of small-scale
impulsive events on the active Sun: Results from IRIS''},
we are using the same IRIS 1400 \ang\ data, extracted
with an automated pattern recognition code during 12
time episodes observed in plage and sunspot regions.
We obtain a total of 23,633 events, quantified in terms
of event areas $A$, radiative energies $E$, and event
durations $T$. We obtain the following results:

\begin{enumerate}
\item{	Fractal dimensions, measured in solar images at
	various wavelengths and spatial resolutions,
	cover a range of $D_A=1-2$, where linear of
	curvi-linear features, as produced by 
	surface flows and magneto-convection,
	characterize the lower limit $D_A \gapprox 1$, 
	while area-filling structures, as they occur
	as a consequence of the chromospheric evaporation 
	process, characterize the upper limit $D_A 
	\lapprox 2$. The mean value $D_A \approx 1.5$
	appears to be a good approximation for SOC models.}

\item{	We calculate a power law fit to the size distribution
	$N(A) \propto A^{-\alpha_A}$ of event areas $A$,
	and find a mean value of $a_A=2.51\pm0.21$ that
	agrees well with the value $a_A=2.33$ expected
	from the theoretical SOC model. Consequently,
	brightenings in plages of the transition region
	are consistent with the generic SOC model.}

\item{	Based on the power law slope $\alpha_A$ we derive
	the fractal dimension $D_A=2/(\alpha_A-1)$, which
	yields a mean observed value of $D_A=1.35\pm0.19$
	and approximately matches the theoretial mean value
	of $D_A=1.5$. Alternatively, we obtain with the
	standard box-counting method an observed value
	of $D_A=1.58\pm0.03$.}

\item{	Synthesizing the measurements of the fractal dimension
	from photospheric, transition region, and coronal data
	we arrive at seven groups that yield the following
	means and standard deviations of their fractal dimension:  

 	\begin{tabular}{ll}
 	photospheric granulation:       & 1.21 $\pm$ 0.07 \\
  	transition region plages: 	& 1.29 $\pm$ 0.15 \\ 
  	transition region sunspots: 	& 1.54 $\pm$ 0.16 \\
  	active region magnetograms: 	& 1.59 $\pm$ 0.08 \\
  	EUV nanoflares: 		& 1.56 $\pm$ 0.08 \\
  	large solar flares: 		& 1.76 $\pm$ 0.14 \\
  	Bastille-Day X5.7-class flare:	& 1.89 $\pm$ 0.05    
 	\end{tabular}					   }				    

\item{	From these seven groups we can discriminate 3 groups
	with significantly different fractal dimensions,
	which implies different physical mechanisms:
	Low values of the fractal dimension $(D_A \approx 1.2-1.3)$
	indicate granulation or transition region plage features, 
	intermediate values of the fractal dimension 
	$(D_A \approx 1.5-1.6)$ indicate sunspots, active 
	region, or nanoflare features, and  
	$(D_A \approx 1.6-1.9)$ indicate large flares. }

\end{enumerate}

The analysis presented here demonstrates that we can distinguish
between (i) physical processes with sparse curvi-linear flows, 
as they occur in granulation, meso-granulation, and super-granulation.
and (ii) physical processes with near space-filling flows, 
as they occur in the chromospheric
evaporation process in solar flares. IRIS data can therefore be used
to diagnose the strength of mass flows in the transition region.
Moreover, reliable measurements of the fractal dimension yields
realistic plasma filling factors that are important in the estimate
of radiative energies and hot plasma emission measures.
Future work on fractal dimensions in multi-wavlength datasets from 
IRIS and AIA/SDO may clarify the dynamics of coronal heating events. 

\acknowledgments
{\sl Acknowledgements:}
We acknowledge constructive and stimulating discussions (in alphabetical order)
with Sandra Chapman, Paul Charbonneau, Henrik Jeldtoft Jensen,
Adam Kowalski, Alexander Milovanov, Leonty Miroshnichenko, 
Jens Juul Rasmussen, Karel Schrijver, Vadim Uritsky, 
Loukas Vlahos, and Nick Watkins.
This work was partially supported by NASA contract NNX11A099G
``Self-organized criticality in solar physics'' and NASA contract
NNG04EA00C of the SDO/AIA instrument to LMSAL.

\clearpage

\def\ref#1{\par\noindent\hangindent1cm {#1}}

\section*{	References	}

\ref{Aschwanden, M.J. and Parnell, C.E. 2002,
        {\sl Nanoflare statistics from first principles: fractal geometry
        and temperature synthesis}, Astrophys. J. 572, 1048}
\ref{Aschwanden, M.J. and Aschwanden P.D. 2008a,
        {\sl Solar flare geometries: I. The area fractal dimension},
        Astrophys. J. 574, 530}
\ref{Aschwanden, M.J. and Aschwanden P.D. 2008b,
        {\sl Solar flare geometries: II. The volume fractal dimension},
        Astrophys. J. 574, 544}
\ref{Aschwanden, M.J. 2011,
        {\sl Self-Organized Criticality in Astrophysics. The Statistics
        of Nonlinear Processes in the Universe}, ISBN 978-3-642-15000-5,
        Springer-Praxis: New York, 416p.}
\ref{Aschwanden, M.J. 2012,
        {\sl A statistical fractal-diffusive avalanche model of a
        slowly-driven self-organized criticality system},
        A\&A 539, A2, (15 p)}
\ref{Aschwanden, M.J. 2014,
        {\sl A macroscopic description of self-organized systems and
        astrophysical applications}, ApJ 782, 54}
\ref{Aschwanden,M.J., Crosby,N., Dimitropoulou,M., Georgoulis,M.K.,
        Hergarten,S., McAteer,J., Milovanov,A., Mineshige,S., Morales,L.,
        Nishizuka,N., Pruessner,G., Sanchez,R., Sharma,S., Strugarek,A.,
        and Uritsky, V. 2016,
        {\sl 25 Years of Self-Organized Criticality: Solar and Astrophysics}
        Space Science Reviews 198, 47-166.}
\ref{Aschwanden, M.J., Scholkmann, F., Bethune, W., Schmutz, W., 
	Abramenko,W., Cheung,M.C.M., Mueller,D., Benz,A.O., 
	Chernov,G., Kritsuk,A.G., Scargle,J.D., Melatos,A., 
	Wagoner,R.V., Trimble,V., Green,W. 2018, 
	{\sl Order out of randomness: Self-organization processes 
	in astrophysics}, Space Science Reviews 214:55}
\ref{Aschwanden, M.J. 2022a,
	{\sl The fractality and size distributions of astrophysical
	self-organized criticality systems},
	ApJ (in press)}
\ref{Aschwanden, M.J. 2022b,
	{\sl Reconciling power-law slopes in solar flare and nanoflare
	size distributions}, ApJL (in press)}
\ref{Balke,A.C., Schrijver, C.J., Zwaan,C., and Tarbell,T.D. 1993,
        {\sl Percolation theory and the geometry of photospheric
        magnetic flux concentrations}, Solar Phys. {\bf 143}, 215.}
\ref{Bak, P., Tang, C., and Wiesenfeld, K. 1987,
        {\sl Self-organized criticality: An explanation of 1/f noise},
        Physical Review Lett. 59(27), 381}
\ref{Bak, P., Tang, C., and Wiesenfeld, K. 1988,
        {\sl Self-organized criticality},
        Physical Rev. A {\bf 38}(1), 364}
\ref{Bak, P. 1996,
        {\sl How Nature Works. The Science of Self-Organized Criticality},
        New York: Copernicus}
\ref{Berrilli, F., Del Moro, D., Russo, S., et al. 2005,
        {\sl Spatial clustering of photospheric structures},
        ApJ 632, 677}
\ref{Bovelet, B. and Wiehr, E. 2001,
        {\sl A new algorithm for pattern recognition and its application
        to granulation and limb faculae}, Solar Phys. {\bf 201}, 13.}
\ref{Cadavid, A.C., Lawrence, J.K., Ruzmaikin, A., and Kayleng-Knight, A.
        1994, {\sl Multifractal models of small-scale magnetic fields},
        Astrophys. J. {\bf 429}, 391.}
\ref{Conlon, P.A., Gallagher, P.T., McAteer, R.T.J., Ireland, J.,
        Young, C.A., Kestener, P., Hewett, R.J., and Maguire, K. 2008,
        {\sl Multifractal properties of evolving active regions}
        Solar Phys. {\bf 248}, 297.}
\ref{De Pontieu, B., Berger, T.E., Schrijver, C.J., and Title, A.M. 1999,
	{\sl Dynamics of transition region 'moss' at high time resolution}.
	Sol.Phys. 190, 419}
\ref{De Pontieu, B. et al.~2014, Sol.Phys. 289, 2733}
\ref{Gallagher, P.T., Phillips, K.J.H., Harra-Murnion, L.K., et al.
        1998, {\sl Properties of the Quiet Sun EUV network},
        A\&A 335, 733}
\ref{Georgoulis, M.K., Rust, D.M., Bernasconi, P.N., and Schmieder, B. 2002,
        {\sl Statistics, morphology, and energetics of Ellerman bombs},
        Astrophys. J. {\bf 575}, 506.}
\ref{Hathaway, D.H., Beck, J.G., Bogart, R.S., et al. 2000,
        {\sl The photospheric convection spectrum}
        SoPh 193, 299}     
\ref{Hirzberger, J., Vazquez, M., Bonet, J.A., Hanslmeier, A., and Sobotka, M.
        1997, {\sl Time series of solar granulation images. I. Differences
        between small and large granules in quiet regions},
        Astrophys. J. {\bf 480}, 406.}
\ref{Ioshpa, B.A., Obridko, V.N., and Rudenchik, E.A. 2008, 
 	{\sl Fractal properties of solar magnetic fields}
 	Astronomy Letters, 34, 210.}
\ref{Janssen, K., Voegler, A., and Kneer, F. 2003,
        {\sl On the fractal dimension of small-scale magnetic structures
        in the Sun}, Astron. Astrophys. {\bf 409}, 1127.}
\ref{Lawrence, J.K. 1991,
        {\sl Diffusion of magnetic flux elements on a fractal geometry},
        Solar Phys. {\bf 135}, 249.}
\ref{Lawrence, J.K. and Schrijver, C.J. 1993,
        {\sl Anomalous diffusion of magnetic elements across the solar
        surface}, ApJ {\bf 411}, 402.}
\ref{Lawrence, J.K., Cadavid, A., and Ruzmaikin, A. 1996,
        {\sl On the multifractal distribution of solar fields},
        ApJ {\bf 465}, 425.}
\ref{Lu, E.T. and Hamilton, R.J. 1991,
 	{\sl Avalanches and the distribution of solar flares}, 
	ApJ 380, L89}
\ref{Mandelbrot, B.B. 1977, {\sl The Fractal Geometry of Nature}.
        W.H.Freeman and Company: New York}
\ref{McAteer, R.T.J., Gallagher, P.T., and Ireland, J. 2005, 
 	{\sl Statistics of Active Region Complexity: A Large-Scale Fractal 
	Dimension Survey}, ApJ 631, 628}
\ref{McAteer,R.T.J., Aschwanden,M.J., Dimitropoulou,M., Georgoulis,M.K.,
        Pruessner, G., Morales, L., Ireland, J., and Abramenko,V. 2016,
        {\sl 25 Years of Self-Organized Criticality: Numerical Detection Methods},
        SSRv 198 217-266.}
\ref{Meunier, N. 1999,
        {\sl Fractal analysis of Michelson Doppler Imager magnetograms:
        a contribution to the study of the formation of solar active regions}.
        Astrophys. J. {\bf 515}, 801.}
\ref{Meunier, N. 2004,
        {\sl Complexity of magnetic structures: flares and cycle phase
        dependence}, Astron. Astrophys. {\bf 420}, 333.}
\ref{Nhalil, N.V., Nelson, C.J., Mathioudakis, M., and Doyle, G.J. 2020,
 	{\sl Power-law energy distributions of small-scale impulsive events 
	on the active Sun: results from IRIS},
 	MNRAS 499, 1385}
\ref{Paniveni, U., Krishan, V., Singh, J., and Srikanth, R. 2005,
        {\sl On the fractal structure of solar supergranulation},
        Solar Phys. {\bf 231}, 1.}
\ref{Paniveni, U., Krishan, V., Singh, J., Srikanth, R. 2010,
        {\sl Activity dependence of solar supergranular fractal dimension},
        MNRAS {\bf 402}(1), 424.}
\ref{Pruessner, G. 2012, {\sl Self-Organised Criticality. Theory, Models
        and Characterisation}, Cambridge University Press: Cambridge.}
\ref{Rieutord, M., Meunier, N., Roudier, T., et al. 2008,
        {\sl Solar supergranulation revealed by granule tracking},
        A\&A 479, L17}
\ref{Rathore, B. and Carlsson, M. 2015, ApJ 811, 80.}
\ref{Rathore, B., Carlsson, M., Leenaarts, J., De Pontieu B. 2015, ApJ 811, 81.}
\ref{Rieutord, M., Roudier, T., Rincon, F., 2010,
        {\sl On the power spectrum of solar surface flows},
        A\&A 512, A4}
\ref{Roudier, T. and Muller, R. 1986,
        {\sl Structure of solar granulation},
        Solar Phys. {\bf 107}, 11.}
\ref{Sharma,A.S., Aschwanden,M.J., Crosby,N.B., Klimas,A.J., Milovanov,A.V.,
        Morales,L., Sanchez,R., and Uritsky,V. 2016,
        {\sl 25 Years of Self-Organized Criticality: Space and Laboratory Plamsas},
        SSRv 198, 167-216.}
\ref{Uritsky, V.M., Paczuski, M., Davila, J.M., and Jones, S.I. 2007,
        {\sl Coexistence of self-organized criticality and intermittent
        turbulence in the solar corona},
        Phys.~Rev.~Lett. 99(2), id. 025001}
\ref{Uritsky, V.M., Davila, J.M., Ofman, L., and Coyner, A.J. 2013,
        {\sl Stochastic coupling of solar photosphere and corona},
        ApJ 769, 62}
\ref{Watkins,N.W., Pruessner, G., Chapman, S.C., Crosby, N.B., and Jensen, H.J.
        {\sl 25 Years of Self-organized Criticality: Concepts and Controversies},
        2016, SSRv 198, 3-44.}

\clearpage

\begin{table}
\begin{center}
\normalsize
\caption{Fractal Dimension obtained from power law fits and from the
box counting method for 12 IRIS datasets.}
\begin{tabular}{lrlll}
\hline
Dataset	  & number of   & power law	& fractal	 & fractal	\\
	  & events      & slope fit	& dimension	 & dimension	\\
	  &		& 		& slope fit      & box counting \\
	  & $n$         & $a_A$	        & $D_{A1}$	 & $D_{A2}$	\\	
\hline
\hline
 	1 &  787 & 2.14 & 1.75 & 1.57 	\\
 	2 & 3119 & 2.32 & 1.52 & 1.57 	\\
 	3 & 2882 & 2.48 & 1.35 & 1.54 	\\
 	4 & 1614 & 2.83 & 1.09 & 1.62 	\\
 	5 & 1106 & 2.67 & 1.20 & 1.62 	\\
 	6 &   65 & 2.47 & 1.36 & 1.61 	\\
 	7 &  118 & 2.37 & 1.45 & 1.60	\\
 	8 & 4412 & 2.50 & 1.33 & 1.55 	\\
 	9 & 4725 & 2.72 & 1.16 & 1.58   \\
       10 & 3064 & 2.28 & 1.56 & 1.55   \\
       11 & 1445 & 2.76 & 1.14 & 1.59 	\\
       12 &  296 & 2.53 & 1.31 & 1.60 	\\
\hline
Mean Obs  &      & 2.51$\pm$0.21 & 1.35$\pm$0.19 & 1.58$\pm$0.03 \\
\hline
Theory    &      & 2.33         & 1.5           &       1.5    \\
\hline
\end{tabular}
\end{center}
\end{table}

\begin{table}
\begin{center}
\normalsize
\caption{Parameters of the standard SOC Model, with fractal dimensions $D_x$
and power law slopes $\alpha_x$ of size distributions).}
\medskip
\begin{tabular}{lll}
\hline
Parameter                    &Power law                                   & Power law  \\
                             &slope                                       & slope      \\
                             &analytical                                  & numerical  \\
\hline
\hline
Euclidean Dimension          &$d =$                                       &3.00        \\
Diffusion type               &$\beta=$                                    &1.00        \\
\hline
Area fractal dimension       &$D_A=d-(3/2)=$                              &1.50=(3/2)  \\
Volume fractal dimension     &$D_V=d-(1/2)=$                              &2.50=(5/2)  \\
\hline
Length                       &$\alpha_L=d=$                               &3.00        \\
Area                         &$\alpha_A=1+(d-1)/D_A=$                     &2.33=(7/3)  \\
Volume                       &$\alpha_V=1+(d-1)/D_V=$                     &1.80=(9/5)  \\
Duration                     &$\alpha_T=1+(d-1) \beta/2=$                 &2.00        \\
Mean flux                    &$\alpha_F=1+(d-1)/(\gamma D_V)=$            &1.80=(9/5)  \\
Peak flux                    &$\alpha_P=1+(d-1)/(\gamma d)=$              &1.67=(5/3)  \\
Spatio-temporal energy       &$\alpha_{E_1}=1+(d-1)/(\gamma D_V+2/\beta)=$ &1.44=(13/9)\\
Thermal energy (h=const)     &$\alpha_{E_2}=1+2/D_A=$                      &2.33=(7/3) \\
Thermal energy (h=A$^{1/2}$) &$\alpha_{E_3}=1+2/D_V=$                      &1.80=(9/5) \\
\hline
\end{tabular}
\end{center}
\end{table}

\begin{table}
\begin{center}
\normalsize
\caption{The fractal dimensions of granules, plages, sunspots, active regions,
nanoflares, and large flares.}
\medskip
\begin{tabular}{llll}
\hline
Phenomenon              &Data, 		&Fractal 		& Reference	\\
			&Method	 	&dimension		& 		\\
			&		& $D_A$			&		\\
\hline
\hline
{\bf Granulation (Photosphere)}	& 	& {\bf 1.21$\pm$0.07} & {\bf Mean} \\
Granules		& 5750 \ang , PA	& 1.25		& Roudier \& Muller (1986) \\	
Granules		& 5257 \ang , PA	& 1.30		& Hirzberger et al. (1997) \\	
Granular cells		& 5257 \ang , PA	& 1.16		& Hirzberger et al. (1997) \\
Granules		& WL, PA		& 1.09		& Bovelet and Wiehr (2001) \\
Granules		& MDI/SOHO, PA		& 1.25		& Paniveni et al. (2005) \\
Supergranulation	& 3934 \ang , Ca II K, PA & 1.23$\pm$0.02 & Paniveni et al. (2010) \\
\hline
{\bf Plages (Transition Region)}& & {\bf 1.26$\pm$0.13} & {\bf Mean} \\  
Plage, TR		& IRIS 1400 \ang	& 1.09		& Nhalil et al. (2020) and this work \\
Plage, TR		& IRIS 1400 \ang	& 1.20		& Nhalil et al. (2020) and this work \\
Plage, TR		& IRIS 1400 \ang	& 1.36		& Nhalil et al. (2020) and this work \\
Plage, TR		& IRIS 1400 \ang	& 1.45		& Nhalil et al. (2020) and this work \\
Plage, TR		& IRIS 1400 \ang	& 1.33		& Nhalil et al. (2020) and this work \\
Plage, TR		& IRIS 1400 \ang	& 1.16		& Nhalil et al. (2020) and this work \\
Plage, TR		& IRIS 1400 \ang	& 1.14		& Nhalil et al. (2020) and this work \\
Plage, TR		& IRIS 1400 \ang	& 1.31		& Nhalil et al. (2020) and this work \\
\hline
{\bf Sunspots (Transition Region)}	&		        & {\bf 1.54$\pm$0.16} & {\bf Mean} \\
Sunspots		& IRIS 1400 \ang	& 1.75		& Nhalil et al. (2020) and this work \\
Sunspot/plage		& IRIS 1400 \ang	& 1.52		& Nhalil et al. (2020) and this work \\
Sunspot/plage		& IRIS 1400 \ang	& 1.35		& Nhalil et al. (2020) and this work \\
Sunspot/plage		& IRIS 1400 \ang	& 1.56		& Nhalil et al. (2020) and this work \\
\hline
{\bf Active regions (Photosphere)}&				& {\bf 1.59$\pm$0.20} & {\bf Mean} \\
Active regions		& BBSO, MG, LA          & 1.56$\pm$0.08 & Lawrence (1991) \\
			& BBSO			&		& Lawrence and Schrijver (1993) \\
Active region plages    & 6302 \ang , Fe I, LA & 1.54$\pm$0.05 & Balke et al. (1993) \\
Active regions		& 7929 \ang , LA	& 1.86$\pm$0.08 & Meunier (1999) \\
Active regions		& 7929 \ang , PA	& 1.58$\pm$0.18 & Meunier (1999) \\
Small scales		& 6302 \ang , Fe I, PA	& 1.41$\pm$0.05 & Janssen et al. (2003) \\
Active regions		& 6768 \ang , Ni I			& 1.80$\pm$0.09 & Meunier (2004) \\
- Cycle minimum		& 6768 \ang , Ni I 	& 1.31$\pm$0.22 & Meunier (2004) \\
- Cycle rise 		& 6768 \ang , Ni I	& 1.80$\pm$0.16 & Meunier (2004) \\
- Cycle maximum		& 6768 \ang , Ni I			& 1.76$\pm$0.04 & Meunier (2004) \\
Active regions		& 6768 \ang , Ni I, BC	& 1.35$\pm$0.30 & McAteer et al. (2005) \\
			& 5250 \ang , Fe I	& 1.5		& Iosnpa et al. (2008) \\
\hline
{\bf EUV nanoflares (Corona)}	&   		& {\bf 1.56$\pm$0.08} & {\bf Mean} \\
nanoflares 		& 171 \ang , EUV, BC	& 1.49$\pm$0.06 & Aschwanden and Parnell (2002) \\
nanoflares              & 195 \ang , EUV, BC	& 1.54$\pm$0.05 & Aschwanden and Parnell (2002) \\
nanoflares   	        & Yohkoh/SXT, AlMg, BC	& 1.65          & Aschwanden and Parnell (2002) \\
\hline
{\bf Large solar flares (Corona)} &		& {\bf 1.76$\pm$0.14} & {\bf Mean} \\	
M-class flares		& 171, 195 \ang , EUV	& 1.62$\pm$0.11 & Aschwanden and Aschwanden 2008a \\
X-class flares		& 171, 195 \ang , EUV	& 1.78$\pm$0.06 & Aschwanden and Aschwanden 2008a \\
Bastille Day flare      & 171, 195 \ang , EUV 	& 1.89$\pm$0.05 & Aschwanden and Aschwanden 2008a \\
\hline
\end{tabular}
\end{center}
\end{table}

\begin{figure}
\centerline{\includegraphics[width=0.9\textwidth]{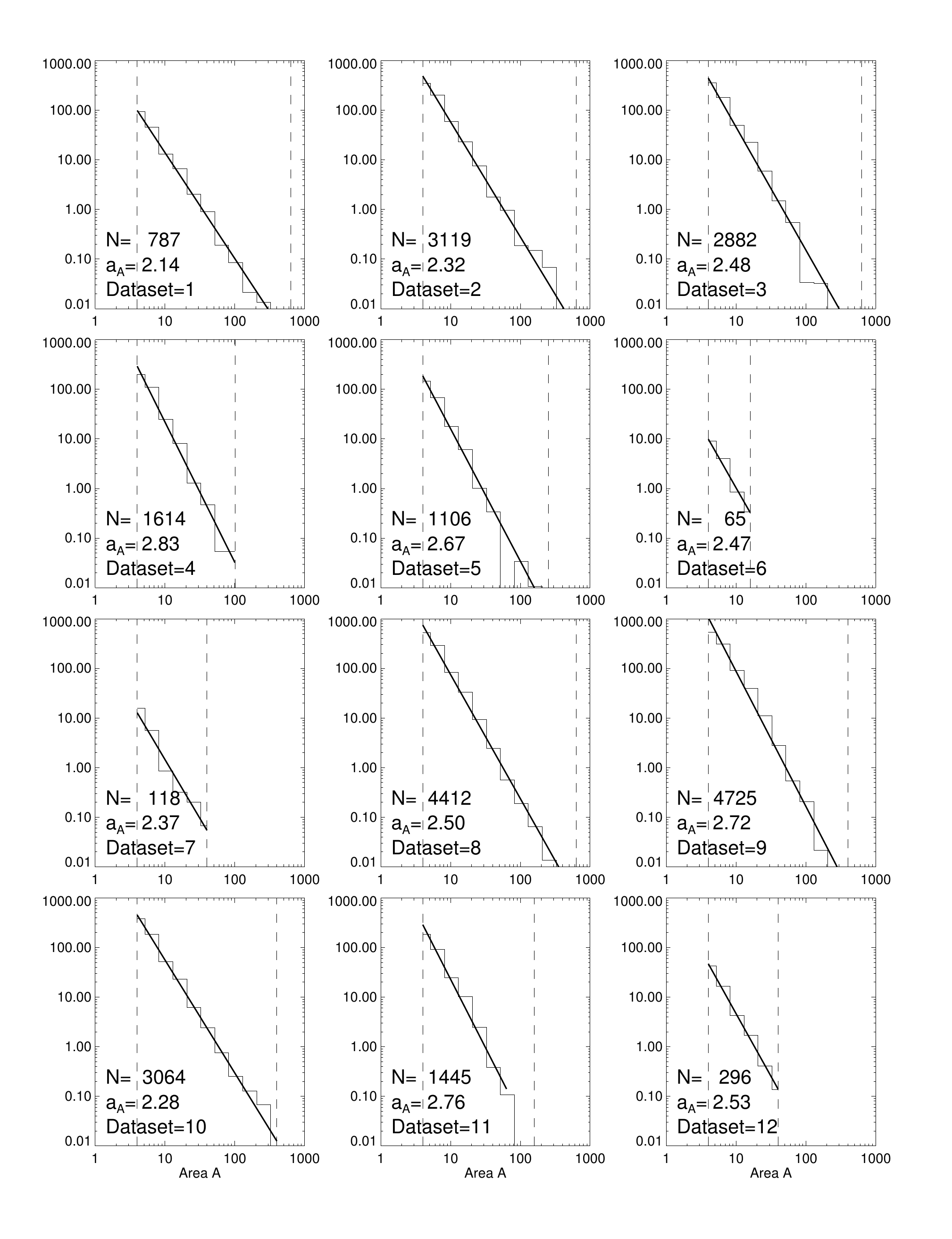}}
\caption{Size distributions of flare areas $A$ for 12 datasets
observed with IRIS SJI 1400 \ang\ in different active regions.}
\end{figure}

\begin{figure}
\centerline{\includegraphics[width=0.9\textwidth]{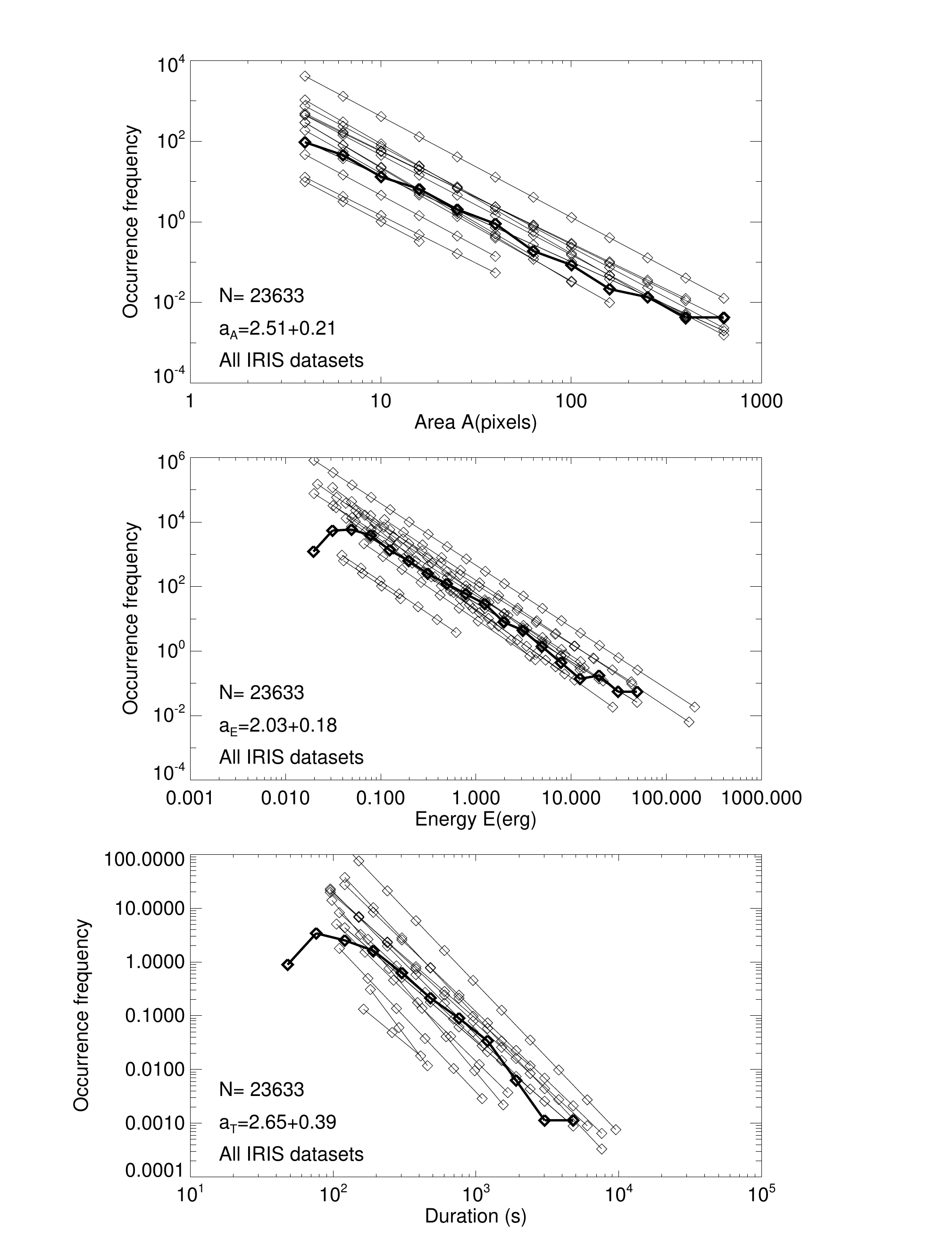}}
\caption{Power law fits to the size distributions of three
SOC parameters: the event area $A$ (top panel), the radiative
energy $E$ (middle panel), and the time duration (bottom panel).
Individual fits to each of the 12 IRIS datasets are indicated
with thin line style, while the fit to all events combined
is indicated with thick line style, and the power law slopes
are given in each panel.}
\end{figure}

\begin{figure}
\centerline{\includegraphics[width=0.9\textwidth]{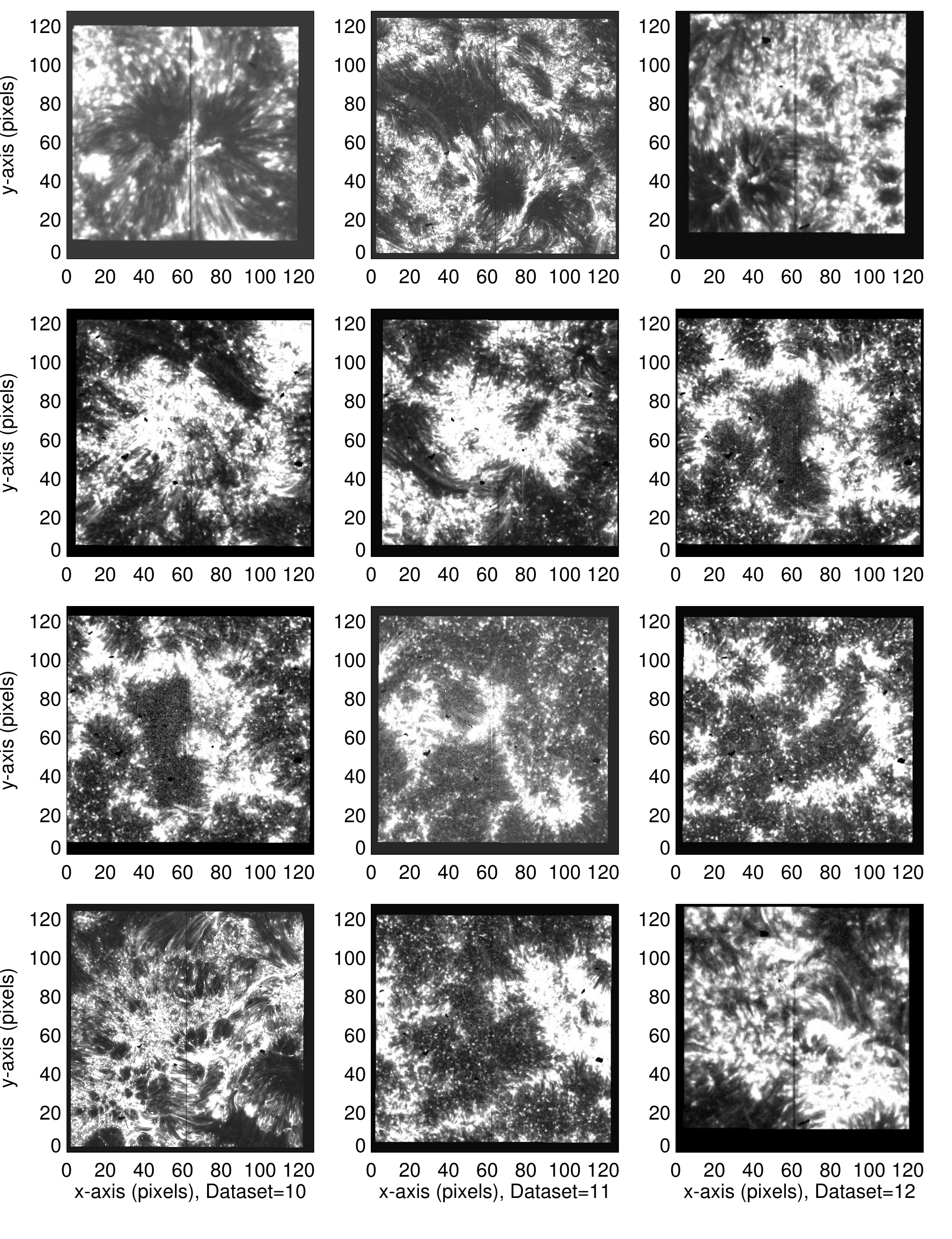}}
\caption{Intensity maps of 12 different active regions,
observed with IRIS SJI 1400 \ang\ .}
\end{figure}

\begin{figure}
\centerline{\includegraphics[width=0.9\textwidth]{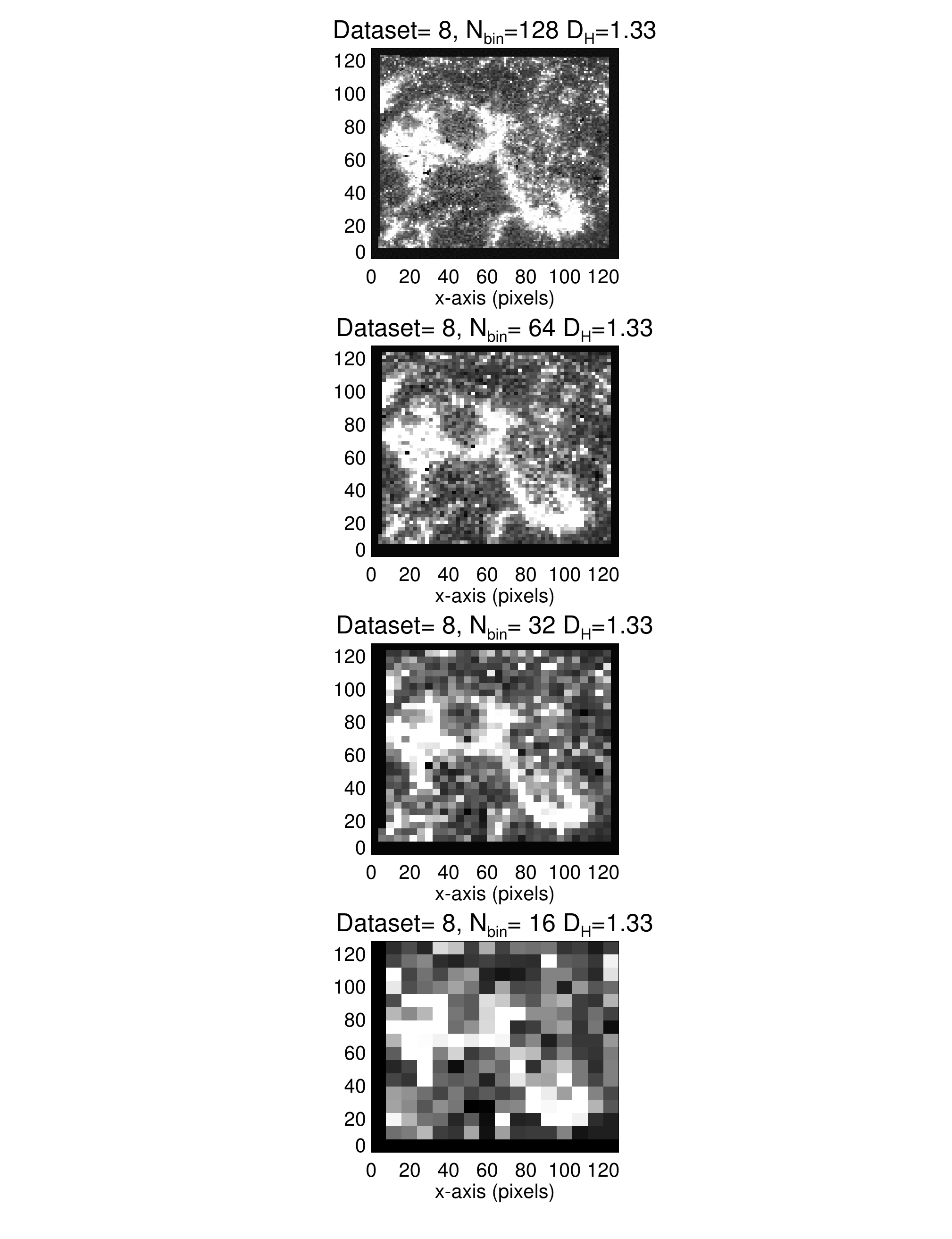}}
\caption{The IRIS dataset 8 is shown with different
spatial resolutions of 128, 64, 32, and 16 bins, which
demonstrates the scale-free definition of the Hausdorff
dimension $D_H=1.33$.}
\end{figure}

\begin{figure}
\centerline{\includegraphics[width=0.9\textwidth]{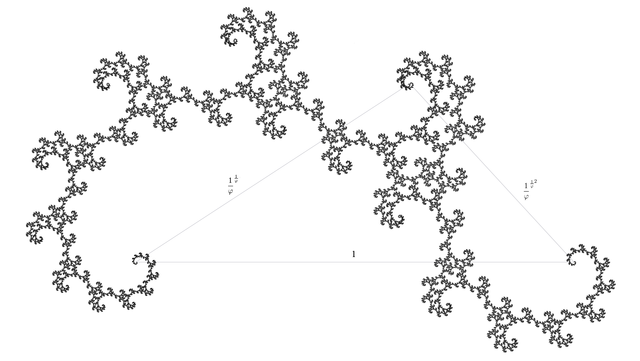}}
\caption{This numerically calculated fractal pattern is called
a {\sl golden dragon} and has a Hausdorff dimension of $D_A=1.61803$.
$(https://en.wikipedia.org/wiki/List\_of\_fractals\_by\_Hausdorff\_dimension)$.
Note the similarity with dataset 8 in Fig.~4}
\end{figure}

\end{document}